\documentclass[epj]{svjour}
\usepackage{graphics}
\begin{document}
\title{$NN$ Correlations Measured in $^3$He(e,e$'$pp)n}
\author{L.B. Weinstein\inst{1} \and Rustam Niyazov\inst{1} for the
CLAS Collaboration
}                     
%
%
\institute{Physics Dept., Old Dominion University, Norfolk, VA 23529,
USA \email{weinstei@physics.odu.edu}}
\date{Received: date / Revised version: date}
%
\abstract{ 
We have measured the $^3$He(e,e$'$pp)n{} reaction in the
Jefferson Lab CLAS with 2.2 and 4.4 GeV electrons.  We looked at the
energy distribution of events with all three nucleons at high momentum
($p > 250$ MeV/c).  This distribution has peaks where two nucleons
each have 20\% or less of the energy transfer (ie: the third or
`leading' nucleon carries most of the kinetic energy).  The angular
distribution of these two `fast' nucleons shows a very large
back-to-back peak, indicating the effect of correlations.
While there is some theoretical disagreement,
experimental evidence, plus calculations at lower energy by
W. Gl\"ockle, indicates that these events are primarily sensitive to $NN$
correlations.  
} 
\maketitle
\section{Introduction}
\label{intro}
The single nucleon energy and momentum distributions in nuclei have
been thoroughly measured by nucleon knockout, pickup and stripping
reactions.  The shapes of these distributions, although not their
magnitudes, are well described by mean-field impulse-approximation
calculations.  The discrepancies between the measured and calculated
magnitudes indicate that nucleon-nucleon correlations are an important
part of the nuclear wavefunction.  To date, there have been almost no
measurements of correlated $NN$ momentum distributions in nuclei.

One signature of correlations is finding two nucleons with large
relative momentum and small total momentum in the initial state.
Unfortunately, the effects of $NN$ correlations are frequently
obscured by the effects of two body currents, such as meson exchange
currents (MEC) and isobar configurations (IC) \cite{janssen00}.  In
order to disentangle these competing effects, a series of
comprehensive measurements are needed.

In order to provide this, we measured
electron scattering from nuclei, A$(e,e^\prime X)$, using the
Jefferson Lab CLAS (CEBAF Large Acceptance Spectrometer), a 4$\pi$
magnetic spectrometer.  The CLAS Multihadron run group comprised of
seven experiments ran in Spring 1999, measuring approximately 500
million events with 1.1, 2.2 and 4.4 GeV polarized electrons incident
on targets from $^3$He to $^{56}$Fe.

This paper will concentrate on the results from  the $^3$He(e,e$'$pp)n{}
reaction which exhibit a strong signature for $NN$ correlations.

\section{The  $^3$He(e,e$'$pp)n{} Measurements}

\begin{figure}[htbp]
  \begin{center}
  \resizebox{0.45\textwidth}{!}{%
    \includegraphics{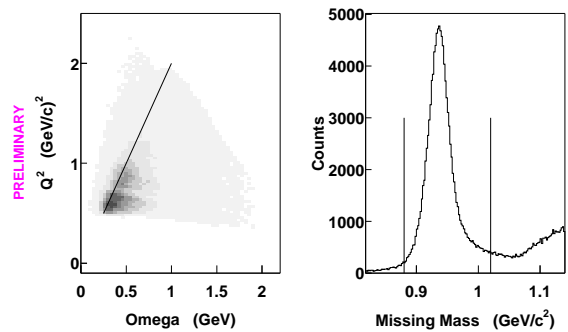}
}
    \caption{a) $Q^2$ vs $\omega$ for $^3$He(e,e$'$pp)n at
    $E_{beam}=2.2$ GeV.  Note the
    huge kinematic acceptance.  b) Missing mass for $^3$He(e,e$'$pp).  
    We cut at the indicated lines to select (e,e$'$pp)n{} events.
      }
	\label{fig:qomega} 
    \end{center}
\end{figure}

We studied electron induced two proton knockout reactions from $^3$He{}
using the CLAS detector and made a cut on the missing mass 
to select $^3$He(e,e$'$pp)n{} events.
Figures \ref{fig:qomega}a and b show the electron acceptance and
undetected neutron missing mass resolution for $E_{beam} = 2.2$
GeV.  The threshold of the CLAS is approximately 250 MeV/c for protons.

Note that all data shown here are {\bf preliminary}.   

Because this is the first time that $^3$He(e,e$'$pp)n{} has been measured
using an almost $4\pi$ detector, our data analysis philosophy 
is to follow and understand the dominant features of the data.  

\begin{figure}[htbp]
  \begin{center} 
  \resizebox{0.45\textwidth}{!}{%
	\includegraphics{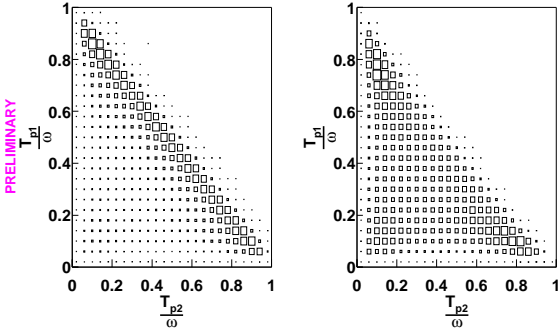} }
	\caption{Nucleon kinetic energy distribution for 2.2 GeV
	$^3$He(e,e$'$pp)n.  The kinetic
	energy of proton 1 divided by $\omega$ is plotted against the
	kinetic energy of proton 2 divided by $\omega$.  The threshold for
	proton detection is $p \ge 250$ MeV/c.  a) all
	events, b) events where $p_n > 250$ MeV/c.  Note the peaks in the
	corners.}  \label{fig:tkin}
	\end{center}
\end{figure}

In order to understand the energy sharing in the reaction, we plotted
the kinetic energy divided by the energy transfer of the first proton
($T_{p1}/\omega$) versus that of the second proton ($T_{p2}/\omega$)
for each event (a lab-frame Dalitz plot).  When we did this, the
dominant feature is a ridge running from the upper left corner (proton
1 has all the energy) to the lower left corner (proton 2 has all the
energy) corresponding to events where the two protons share the energy
transfer and the neutron is a low momentum `spectator' (see Figure
\ref{fig:tkin}a).  When we cut on this ridge, we see that the opening
angle of the two protons has a large peak at 90$^o$, indicating that
it is due primarily to hard final state rescattering (i.e.: photon
absorption on one proton followed by billiard ball rescattering on the
second proton).

Since we are not interested in final state rescattering, we eliminated
those events and focussed on events where all three nucleons have
momentum greater than 250 MeV/c (Figure \ref{fig:tkin}b).  In this
case we see three peaks at the three corners of the plot,
corresponding to events where two `fast' nucleons each have less than
20\% of the energy transfer and the third `leading' nucleon has the
remainder.  We call the two nucleons `fast' because $p \gg p_{fermi}$.
These peaks are much more pronounced at $E_{beam} = 4.4$ GeV (not
shown).  We cut on these peaks where the two fast nucleons each have
less than 20\% of the energy transfer and where all three nucleons
have $p> 250$ MeV/c.

\begin{figure}[htbp]
  \begin{center}
  \resizebox{0.45\textwidth}{!}{%
	\includegraphics{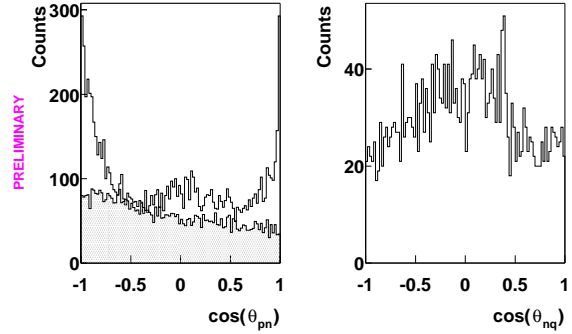} }
    \caption{a) Opening angle of the fast $pn$ pairs for events in the
    upper left and lower right corners of Fig. \ref{fig:tkin}-b.  The
    backward-peaked histogram shows the data, the filled histogram
    shows the results of a fire ball phase space simulation assuming
    three body absorption of the virtual photon and phase space decay
    (with arbitrary normalization).  b) The angle between the neutron
    in the fast $pn$ pair and $\vec q$ where $p_{\perp} < 300$ MeV/c.
    }
\label{fig:openang} 
\end{center}
\end{figure}

\begin{figure}[!p]
  \begin{center}
  \resizebox{0.5\textwidth}{!}{\rotatebox{-90}{%
    \includegraphics{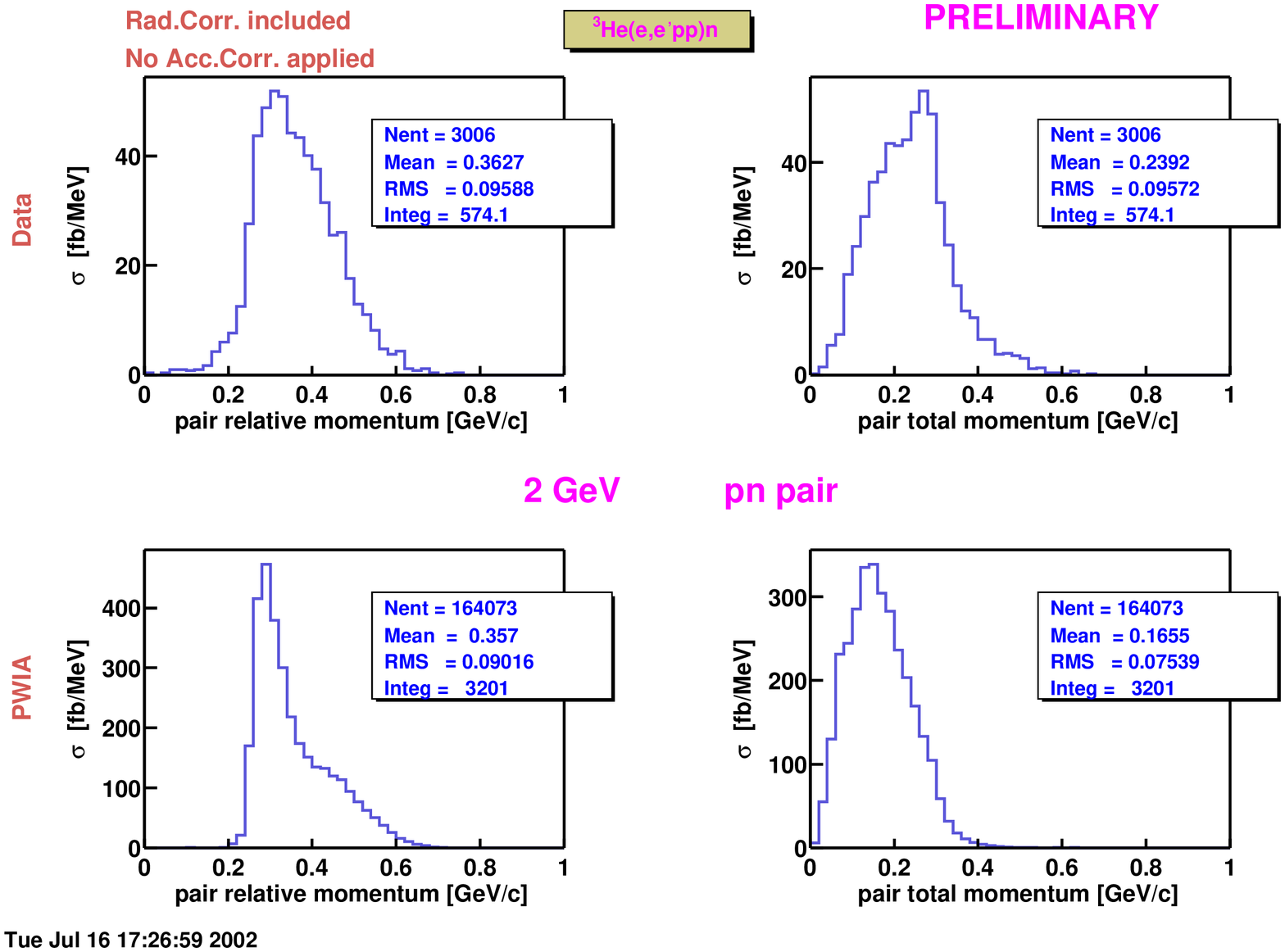}
}}
    \caption{Cross section for events with a leading proton and a fast
    $pn$ pair at $E_{beam}=2.2$ GeV.  a) Data: relative momentum;
    b) Data: total momentum; c) PWIA: relative momentum;
    d) PWIA: total momentum.
    }
\label{fig:crosssec} 
\end{center}
\end{figure}

Then we looked at the opening angle of the two fast nucleons.  Figure
\ref{fig:openang}-a shows the pair opening angle for fast $pn$ pairs
with a leading proton.  Note the large peak at 180 degrees
($\cos\theta_{NN} \approx -1$).  The distribution for fast $pp$ pairs
with a leading proton is identical.  The peak is not due to the cuts,
since we do not see it in a fire ball phase space simulation assuming
three body absorption of the virtual photon and phase space decay.  It
is also not due to the CLAS acceptance since we see it both for leading
protons (which we detect) and leading neutrons (which we infer from
missing mass).  This back-to-back peak is a strong indication of
correlated $NN$ pairs.

\section{Studying Correlated Pairs}

Now consider these presumably correlated pairs.  Since we
believe that we have observed events where the leading nucleon
absorbed the virtual photon and the two fast nucleons are emitted back
to back, we  cut on the perpendicular momentum of the leading
nucleon to deemphasize rescattering   ($p_\perp <
300$ MeV/c).  This cut selects the back-to-back
events very cleanly.  Unfortunately, there are only 3400 fast $pn$ and
1100 fast $pp$ events remaining in the entire 2.2 GeV data set (and
ten times fewer at 4.4 GeV).

If the fast back-to-back $NN$ pairs are really uninvolved in the
photon absorption, then they should be distributed isotropically.  You
can see this in  the angular distribution of the
neutrons with respect to $\vec q$ (see Figure \ref{fig:openang}-b).
Further evidence that the fast $NN$ pair is uninvolved in absorbing
the virtual photon comes from the average momentum of the pair along
$\vec q$.  This is about 0.07 GeV/c for $E_{beam} = 2.2$ GeV and about
0.1 GeV/c for $E_{beam} = 4.4$ GeV, much less than the average
momentum transfers of $Q^2 = 0.7$ and 1.4 (GeV/c)$^2$ respectively.

The fast $NN$ pair 
relative ($p_{rel} = {1\over 2}\vert\vec p_1 - \vec p_2\vert$) and
total ($p_{total} = \vert\vec p_1 + \vec p_2\vert$) 
momentum distributions are shown in Figures \ref{fig:crosssec}a) and
b) for  fast $pn$  pairs at
$E_{beam} = 2.2$ GeV.   The distributions (not shown) are very
similar for both $pn$ and $pp$ pairs at both $E_{beam} = 2.2$ and 4.4 GeV.

Thus, because when we select a quasifree leading nucleon the fast $NN$
pairs are:
\begin{itemize}
\item Back to Back,
\item Isotropic and
\item Have small average momentum along $\vec q$
\end{itemize}
we conclude that the fast $NN$ pair is not involved in absorbing the
virtual photon.  Because we measure similar total and relative
momentum distributions for
\begin{itemize}
\item $pp$ and $pn$ pairs and
\item $0.5 < Q^2 < 1$ ($E_{beam} = 2.2$ GeV) and $1 < Q^2 < 2$
(GeV/c)$^2$  ($E_{beam} = 4.4$ GeV) 
\end{itemize}
we conclude that we have measured bound state $NN$ correlations.

We appear to have measured $NN$ correlations in $^3$He by
striking the {\bf third} nucleon and detecting the correlated pair.
This is similar to other proposed correlation searches where you
strike one nucleon of a correlated pair and detect the other nucleon
leaving the nucleus.  However, these other searches suffer from the weakness
that their proposed signal can also be due to two body currents (eg:
photon absorption on an exchanged meson).

\section{Comparison to Theory}

Calculations by W. Gl\"ockle \cite{glockle} at lower energy strengthen this
conclusion.  He calculated the  $^3$He(e,e$'$pp)n{}  cross section where
the leading nucleon has momentum $\vec p_N = \vec q$ and the other two
nucleons have total momentum $p_{total} = 0$ for various values of the
momentum transfer, $400 \le \vert \vec q \vert \le 600$ MeV/c, and
relative momentum.  He found that
\begin{enumerate}
\item MEC did not contribute, 
\item rescattering of the leading nucleon did not contribute, and
\item the continuum state interaction of the outgoing $NN$ pair
decreased the cross section by a factor of approximately 10 relative
to the PWIA result.
\end{enumerate}
Thus, he found that this reaction is a very clean way to measure the
overlap integral between the $NN$ continuum state and the same two
nucleons in the bound state.  

We compared our results to three other calculations, 1) a Plane Wave
Impulse Approximation (PWIA) calculation by M. Sargsian \cite{misak}
using Gl\"ockle's bound state wave function with no final state
interactions, 2) a calculation by J.-M. Laget \cite{laget} using a
Faddeev wave function from P. Sauer and including one-, two-, and
three- body mechanisms as well as rescattering terms, and 3) a
home-made model of pion production on the struck proton followed by
pion absorption on the remaining $pn$ pair.  We averaged all of the
models over the CLAS acceptances and cuts using a monte carlo.

The pion production and rescattering model used pion production cross
sections from the MAID parametrization \cite{maid}, pion absorption on
deuterium from the SAID parametrization \cite{said}, and proton
initial momentum distributions in $^3$He{} from  (e,e$'$p)
measurements \cite{jans}.  This model failed in several key respects.
While it did produce a large back-to-back peak in the $NN$ angular
distribution (since a soft pion transfers a lot of energy but very
little momentum), a) the average energy transfer was much larger than
the data (typical of the $\Delta(1232)$), b) the relative momentum
distribution was too large (since the minimum relative energy $E_{rel}
= m_\pi$), and c) the ratio of the number of fast $pn$ pairs to fast
$pp$ pairs was much lower than the data (1 instead of 3).  Thus, while
this mechanism might be very important for other three nucleon
knockout experiments, it does not explain this data.

Preliminary calculations from Laget describe the kinetic energy,
relative momentum and total momentum distributions very well, both
qualitatively and quantitatively.  They indicate that one body
knockout plus rescattering cannot describe the data and that
three-body mechanisms are needed.  However, the virtual photon
distribution in this calculation is significantly different from the
data (peaked in the delta region rather than the quasielastic),
indicating a different reaction mechanism.

The PWIA calculation of Sargsian has $Q^2$ vs $\omega$, $NN$ pair
opening angle, and relative and total momentum  distributions that
are consistent with the data (see Figure \ref{fig:crosssec}c--d).  It is
a factor of 6 larger than the data which is consistent with the
expected effects of the $NN$ continuum state interaction calculated by
Gl\"ockle.  However, it predicts 5 $pn$ pairs for each $pp$ pair
versus 3 in the data and it predicts a ratio of 4 for
$\sigma(E_{beam}=2.2) / \sigma(E_{beam}=4.4)$ versus 11 for the data.

More calculations are clearly needed to resolve these discrepancies.

\section{Summary}
We have studied the $^3$He(e,e$'$pp)n{} reaction, selecting events where one
nucleon has most of the kinetic energy and has less than 300 MeV/c of
momentum perpendicular to $\vec q$.  When we do this, we see
isotropic, back-to-back, fast $NN$ pairs with small average momentum
along $\vec q$.  We have measured the total and relative momentum
distributions of these pairs and found that they do not depend
significantly on isospin ($pp$ vs $pn$ pairs) or on momentum transfer.

PWIA calculations reproduce many features of the data.  Calculations
by Gl\"ockle at lower energy indicate that the cross section depends
primarily on the overlap integral between the continuum state and
bound state of the $NN$ pair.  Neither meson exchange currents nor the
final state rescattering of the leading nucleon appear to contribute
to the cross section.  However, calculations by Laget indicate that
three body mechanisms are required.  More theoretical work is needed
to resolve these issues.

Thus, by measuring $^3$He(e,e$'$pp)n, we might have directly
measured $NN$ correlations  without any significant contamination
from other processes by striking the {\bf third} nucleon and detecting
the spectator correlated pair.

\section*{Acknowledgments}
I thank Jean-Marc Laget and Misak Sargsian for their calculations and
physics insight.  This work was supported by a grant from the US
Department of Energy.


\begin{thebibliography}{99}
\bibitem{janssen00} S. Janssen {\sl et al.}, Nucl. Phys. \textbf{A672},
 (2000) 285.
\bibitem{glockle} W. Gl\"ockle, proceedings of the ``5$^{th}$ Workshop
on Electromagnetically Induced Two-Hadron Emission'', Lund, Sweden, 2001.
\bibitem{misak} M. Sargsian, private communication.
\bibitem{laget} G. Audit {\it et alia}, Nucl. Phys. \textbf{A614},
(1997) 461; J.M. Laget, J. Phys. G \textbf{14}, (1988) 1445.
\bibitem{maid} D. Drechsel, O. Hanstein, S.S. Kamalov, L. Tiator,
Nucl. Phys. \textbf{A645} (1999) 145--174; nucl-th/9807001.  Available at
http://www.kph.uni-mainz.de/MAID/maid2000/maid2000.html.
\bibitem{said} C.H. Oh, R.A. Arndt, I.I. Strakovsky, R.L. Workman,
Phys. Rev. C {\bf 56}, 635 (1997).
\bibitem{jans} E. Jans, {\it et alia}, Nucl. Phys. \textbf{A475}, (1987) 687.


\end{thebibliography}
\end{document}